\documentclass[%
 reprint,
 amsmath,amssymb,
 aps,
pra,
]{revtex4-1}

\usepackage{graphicx}
\usepackage{dcolumn}
\usepackage{bm}

\usepackage{amsfonts,latexsym,fancyhdr,graphicx,bbold}

\DeclareMathOperator{\Tr}{Tr}

\newcommand{\ket}[1]{\lvert #1 \rangle}								
\newcommand{\bra}[1]{\langle #1 \rvert}               
\newcommand\bigzero{\makebox(0,0){\text{\huge0}}} 

\begin{document}

\preprint{APS/123-QED}

\title{High fidelity GHZ generation within nearby nodes}

\author{Valentina Caprara Vivoli}
\email{capraravalentina@gmail.com}
 \affiliation{QuTech, Delft University of Technology, Lorentzweg 1,
   2628 CJ Delft, The Netherlands}

\author{J\'er\'emy Ribeiro}
\affiliation{QuTech, Delft University of Technology, Lorentzweg 1, 2628 CJ Delft, The Netherlands}

\author{Stephanie Wehner}
\affiliation{QuTech, Delft University of Technology, Lorentzweg 1, 2628 CJ Delft, The Netherlands}


\begin{abstract}
Generating entanglement in a distributed scenario is a fundamental
task for implementing the quantum network of the future.
We here report a protocol that uses only linear optics for
generating GHZ states with high fidelities in a nearby
node configuration. Moreover, we analytically show that the scheme is optimal for certain initial states in providing the highest success probability, and, then, the highest
generation rate for sequential protocols. Finally, we give some estimates for the generation rate in a
real scenario.
\end{abstract}

\maketitle

\section{Introduction}
\begin{figure}[h]
 \begin{center}
 \includegraphics[width=190pt]{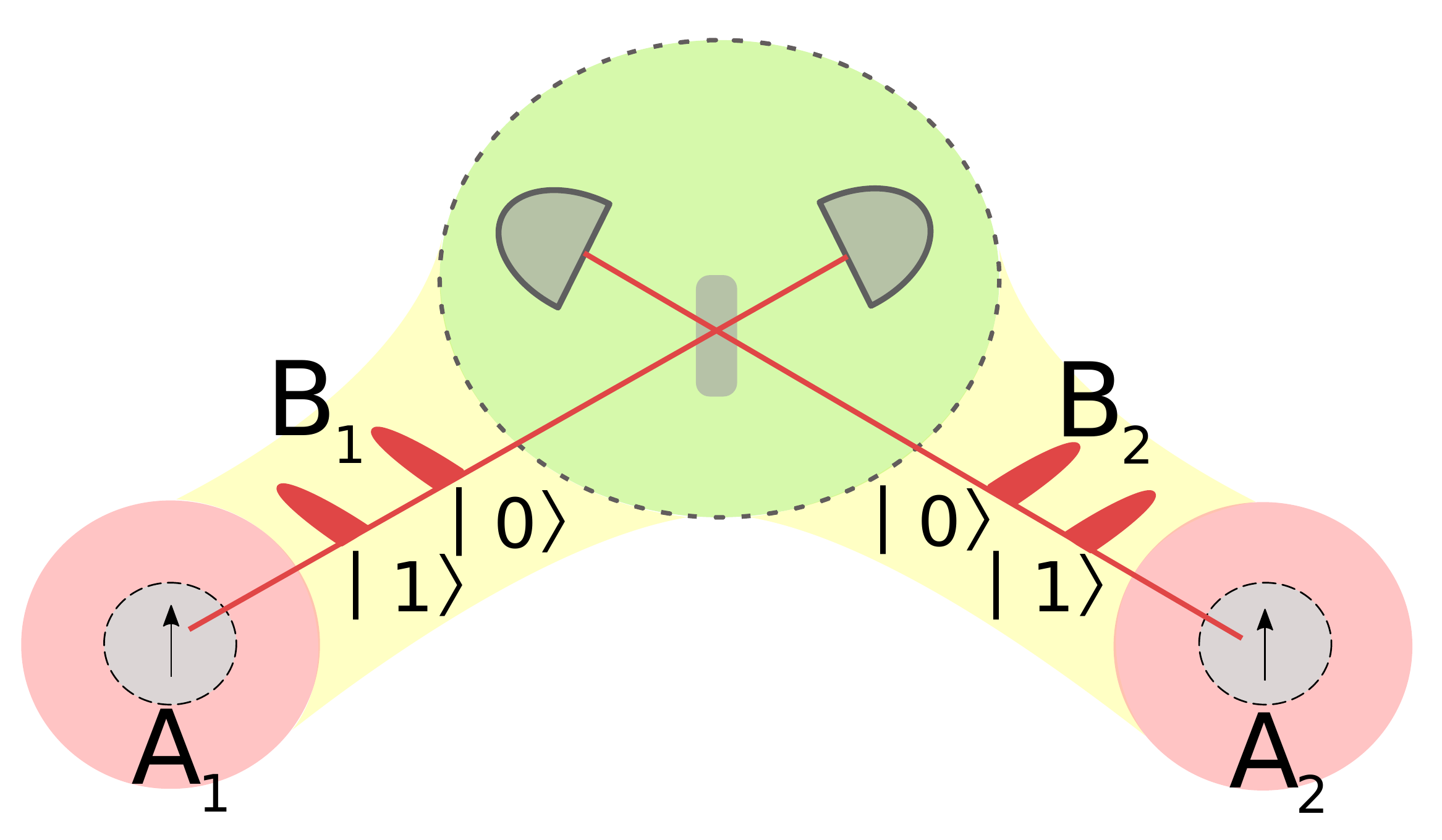}
 \end{center}
 \caption{Barrett-Kok scheme \cite{Barrett05, Lim05}. Two nodes constituted by
   two-level systems are optically excited so that they generate
   matter-photon entanglement,
   i.e. $(\ket{00}+\ket{11})_{A_jB_j}$. The photons are sent to a
   common station where a partial Bell measurement, that can distinguish only two Bell states, is performed. \label{BarrettKok}}
 \end{figure}

\hspace{.5cm}
Entanglement has revealed several interesting applications in
quantum networks. For example, bipartite entanglement can be used for
quantum cryptography tasks, i.e. quantum key distribution (QKD)
\cite{Ekert91, Pironio09}, teleportation \cite{Bennett93}, superdense
coding \cite{Bennett92}, and bit commitment \cite{Lo97,Aharon16}. However, more and more interest has been
recently put in the study of multipartite entanglement. Several uses are
nowadays known, such as, for example, reducing communication
complexity \cite{Buhrman01,Buhrman10}, and
distributed quantum computation \cite{Cleve97, Grover97,Li15,Li16}.
Furthermore, there are multiple uses in quantum cryptography, namely quantum secret
sharing \cite{Hillery99}, N-partite quantum key distribution, also
known as conference key agreement
\cite{Epping16}, and anonymous transfer
\cite{Christandl05}. Multipartite entanglement could also be extremely
useful for implementing quantum repeaters of second and third
generation \cite{Muralidharan16,Gottesman99,Muralidharan14,Munro12}. Finally, it has recently been pointed out that the use of
multipartite entanglement could be fruitful for synchronizing several
atomic clocks \cite{Komar14}.
GHZ states \cite{GHZ89} are particularly suitable for all these purposes. It is, thus, an interesting question how we can best generate such state in a distributed scenario, i.e. where the qubits between which the entanglement is shared can interact only through ancillary modes.
In the case of two-qubit entanglement, there
is already a protocol \cite{Barrett05, Lim05} (see Fig. \ref{BarrettKok}), using ancillary photonic modes, that works pretty well for generating maximally
entangled states in matter systems and low loss regimes. However,
it is still not very clear how this scheme can be extended in the case
of multipartite entanglement.
In the latter case, there have been some proposals as well \cite{Komar16,Nickerson14,Bose98,Zeilinger97,Cuquet12}. They all
consist of two steps: i. Maximally entangled states are generated
between two nodes through Bell measurements, ii. local probabilistic
operations inside the nodes or additional Bell measurements are
realized, generating multipartite entanglement all along the network.
Concerning the network structure, it can vary from long chains of nodes to
closed configurations with nearby nodes. Even though the first
structure allows to cover long distances, the second gives the
possibility to make interact all the nodes between them through a
central station. Little effort has been done so far in the study of
fidelity and generation rate, and for the most without any study of
the trade-off between these two quantities.\newline

\noindent {\bf Remark:} From now on we will loosely use the term Bell measurement to refer to the measurement performed by interference of two optical modes on a beam splitter followed by two non-photon number resolving detectors, where the detection of a photon on one detector determines the success.

\subsection*{Results and structure of the paper}

The goal of this work is to study optimal ways of generating
N-partite GHZ states between nearby nodes with very high fidelity, with realistic settings,
i.e. in the presence of noise. To do so, we focus on protocols that
reduce the number of steps when decoherence is involved by avoiding the use of local operations: These protocols only use binary measurements
that output a flag with values {\it success} or {\it failure}.
More precisely each of the $N$ data qubits (subsystem $A$) will be entangled to an ancillary
qubit (the $N$ ancillary qubits that form subsystem $B$) that is sent to a ``Central Station''
where a joint binary Positive-Operator Valued Measurement (POVM)  will
be performed on the $N$ ancillary qubits outputting either {\it success} or {\it failure} (see Fig.~\ref{FigureProof}).
We, then, optimize over all binary POVMs (performed by the ``Central Station'') and over
all states formed by the data qubits and the ancillaries.
\begin{itemize}
  \item In section \ref{1}, we work in a noise free model, where
   we first show an upper-bound on the product $F\cdot p_{\text{succ}}$ as a function
    of the initial state,
    where $F$ is the fidelity between the $N$-partite GHZ state and the final state
    conditioned on {\it success}, and $p_{\text{succ}}$ is the success probability.\\
    We then explicitly show that there exists
    a binary POVM that saturates the upper-bound for $F \cdot p_{\text{succ}}$. \\
    Finally we search for initial
    states that allow to get a fidelity $F=1$ between the final output state and the GHZ state.\\
    We conclude that there exists a measurement (determined by the projector onto the $N$-GHZ state)
    that allows for creation of an $N$-partite
    GHZ state with an optimal success probability of $2^{-N}$.
    \item In section \ref{Sec:Implementation}, we show how to implement
    the above mentioned optimal POVM with only linear optics and non-photon resolving
    detectors (see Fig.~\ref{OpticalSys}). It turns out
    that to perform this measurement we only need measurements between each two consecutive nodes.
    It means that a ``Central Station'' is not needed. This allows for more flexibility
    in the implementation of the measurement, which can be used to reduce losses and other sources of noise.
    This implementation is inspired by an old work \cite{Zukowski95} in all photonic systems,
    and we show that it is a natural extension of the scheme proposed in \cite{Barrett05,Lim05}.
    \item Finally in section \ref{Sec:Perf}, we give some results in a scenario that could be reasonable in
    the near future. We focus on the
    entanglement generation rate, comparing it for different numbers of
    nodes and inter-node distance.
\end{itemize}

\section{Nodes-Center Scenario\label{1}}
\begin{figure}[h]
 \begin{center}
\includegraphics[width=250pt]{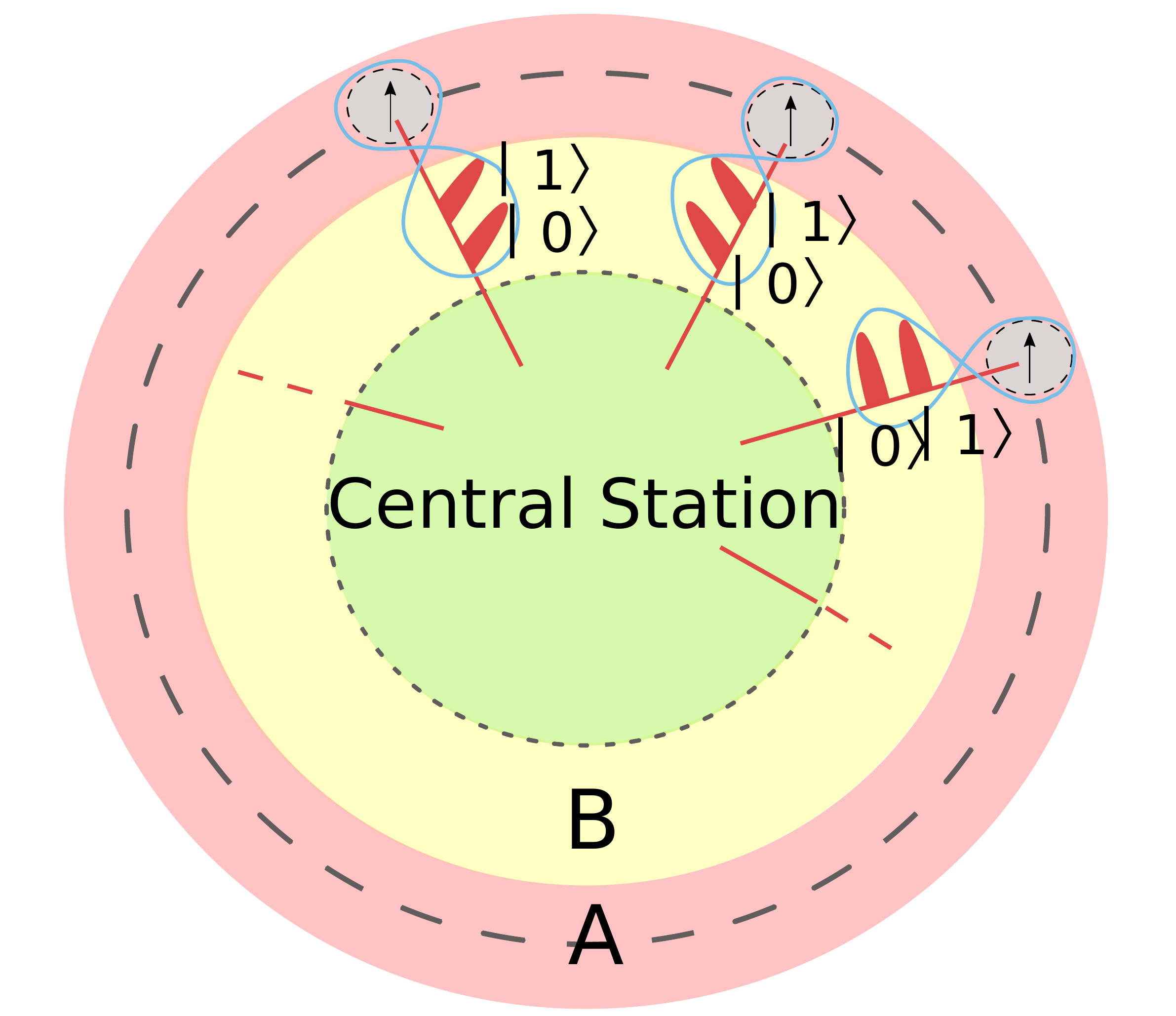}
 \end{center}
 \caption{Nodes-Center scenario. The entire system is composed by two
   subsystems A (pink shell) and B (yellow shell). Each subsystem is composed by N qubits. The
   qubits between A and B are entangled in pairs, i.e. there are N
   entangled pairs $(\sqrt{1-\epsilon}\ket{00}+\sqrt{\epsilon}\ket{11}) _{A_j
  B_j}$. The N qubits of subsystem B are analyzed together through a
POVM in a central station.
  \label{FigureProof}}
 \end{figure}

\hspace{.5cm}
In this section we firstly show that there is an upper-bound for the
product of the fidelity ($F$) between the GHZ state and the final
state, and the success probability ($p_{\text{succ}}$) of the POVM, depending on the initial state. Secondly, we derive the map that allows to reach
the upperbound and show that only for $p_{\text{succ}}=2^{-N}$ it is possible
to saturate the upperbound and get $F=1$.
 In order to do so, let's consider the scenario represented in
Fig. \ref{FigureProof}.
The total system is composed by two subsystems A and B, each one composed by N qubits. We take the
initial state to be
\begin{equation}
\ket{\Psi_{\text{in}}}_{AB}=\bigotimes^N_{j=1}(\sqrt{1-\epsilon}\ket{00}+\sqrt{\epsilon}\ket{11}) _{A_j
  B_j},
\end{equation}
where $A_j$ ($B_j$) are qubits, and $0\le \epsilon\le 1$.
We assume that in the central station it is possible to perform an
arbitrary POVM.

\subsection{Optimal $F\cdot p_{\text{succ}}$}
\hspace{.5cm}
Our goal, here, is to derive an upper-bound for $F\cdot
p_{\text{succ}}=\Tr((\ket{\text{GHZ}}\bra{\text{GHZ}}_A\otimes\Pi_B^{\text{succ}})\cdot\ket{\Psi_{\text{in}}}\bra{\Psi_{\text{in}}}_{AB})$,
when optimizing our POVM elements $\Pi_B^{\text{succ}}$ on $B$ indicating successful generation. Let's consider the following series of inequalities,
\begin{equation}\begin{split}\label{eqF'}
F\cdot p_{\text{succ}}&\le F\cdot p_{\text{succ}}+F_{\text{fail}}\cdot
p_{\text{fail}}=\\
&=\Tr(\ket{\text{GHZ}}\bra{\text{GHZ}}_A\otimes\Pi^{\text{succ}}_B\cdot\ket{\Psi_{\text{in}}}\bra{\Psi_{\text{in}}}_{AB})\\
&+\Tr(\ket{\text{GHZ}}\bra{\text{GHZ}}_A\otimes (\mathbb{1}-\Pi^{\text{succ}})_B\cdot\ket{\Psi_{\text{in}}}\bra{\Psi_{\text{in}}}_{AB})\\
&=F\left(\Tr_B(\ket{\Psi_{\text{in}}}\bra{\Psi_{\text{in}}}_{AB}),\ket{\text{GHZ}}_A\right)\\
&=\frac{1}{2}\left((1-\epsilon)^N+\epsilon^N\right).
\end{split}\end{equation}
Here, $p_{\text{fail}}$ is the
probability that the measurement does
not succeed, and $F_{\text{fail}}$ is the overlap between the
GHZ state and the state that would result from the fail outcome.
$F\left(\Tr_B(\ket{\Psi_{\text{in}}}\bra{\Psi_{\text{in}}}_{AB}),\ket{\text{GHZ}}_A\right)=F(\Psi_{\text{in}}^A;\text{GHZ})$
is the fidelity between the initial state in A and the GHZ state.
The previous upperbound can be interpreted saying that the maximal
amount of entanglement that can be extracted from subsystem A does not
depend on subsystem B. In the case when several success events are considered, the proof follows the same procedure for upperbounding the sum $\sum_i F^i\cdot p^i_{\text{succ}}$. One finds a sum of terms of the same form of the one of the fourth line of Eq. \eqref{eqF'}, where instead of the GHZ state there are several different GHZ-like states.

\subsection{Optimal map $\Pi_B^{\text{succ, opt}}$}
\hspace{.5cm}
For $\epsilon$ very small, the bound in Eq. \eqref{eqF'} is close to $\frac{1}{2}$. We now ask
\begin{enumerate}
\item whether this upper-bound is attainable, and
\item what is the maximal fidelity in this case.
\end{enumerate}
In order to answer to these questions, it is necessary to find the
POVM that allows us to reach $F(\Psi_{\text{in}}^A;\text{GHZ})$.  Suppose that the bound \eqref{eqF'} is attainable by
$F^{\text{opt}}\cdot p_{\text{succ}}^{\text{opt}}$ and we look for the
element $\Pi_B^{\text{succ, opt}}$ of a POVM
s.t.
\begin{equation}
p_{\text{succ}}^{\text{opt}}=\Tr_B\left(\Pi_B^{\text{succ, opt}}\cdot\left[(1-\epsilon)\ket{0}\bra{0}+\epsilon\ket{1}\bra{1}\right]_B^{\otimes  N}\right)
\end{equation}
is minimal and, hence, $F^{\text{opt}}$ is maximal.
$\Pi_B^{\text{succ, opt}}$ can be written
as a $2^N\times 2^N$ square
matrix of elements $e_{lm}$.
One has, then,
\begin{equation}
p_{\text{succ}}^{\text{opt}}=e_{11}(1-\epsilon)^N+e_{2^N  2^N}\epsilon^N,\label{P}
\end{equation}
and
\begin{equation}\begin{split}\label{FP}
F^{\text{opt}}\cdot
p_{\text{succ}}^{\text{opt}}=\frac{1}{2}&\bigg[e_{11}(1-\epsilon)^N+e_{2^N
    2^N}\epsilon^N\\
&+(e_{12^N}+e_{2^N 1})\sqrt{\epsilon(1-\epsilon)}^N\bigg]\\
=&\frac{1}{2}\left[p_{\text{succ}}^{\text{opt}}+(e_{12^N}+e_{2^N 1})\sqrt{\epsilon(1-\epsilon)}^N\right].
\end{split}\end{equation}
The minimization of $p_{\text{succ}}^{\text{opt}}$ is subjected to 5 conditions. The first condition
derives from the bound \eqref{eqF'}, i.e.
\begin{enumerate}
\item

\begin{align*}
 F^{\text{opt}}\cdot p_{\text{succ}}^{\text{opt}}&\\
=&\frac{1}{2}\Tr_B\bigg(\Pi_B^{\text{succ, opt}}\left(\sqrt{1-\epsilon}^N\ket{0}^{\otimes
N}+\sqrt{\epsilon}^N\ket{1}^{\otimes N}\right)\\
&\,\,\,\,\,\,\,\,\,\,\,\,\,\,\,\,\, \left(\sqrt{1-\epsilon}^N\bra{0}^{\otimes
N}+\sqrt{\epsilon}^N\bra{1}^{\otimes N}\right)_B \bigg)\\
=&\frac{1}{2}\left((1-\epsilon)^N+\epsilon^N\right) .
\end{align*}
\end{enumerate}
From the fact that $\Pi_B^{\text{succ, opt}}$ is an element of a POVM, we can derive the condition $0\le\Pi_B^{\text{succ, opt}}\le \mathbb{1}$. This leads us to the following four necessary conditions:
\begin{itemize}
\item[2.] $0\le e_{11}\le 1$,
\item[3.] $0\le e_{2^N 2^N}\le 1$,
\item[4.] $e_{1 2^N}=e^*_{2^N 1}$,
\item[5.] $e_{1 2^N}e_{2^N 1}\le \text{min}(e_{11}e_{2^N 2^N},(1-e_{11})(1-e_{2^N 2^N}))$.
\end{itemize}
All the elements $e_{lm}$ with $l,m \not=1,2^N$ do not influence the
values of Eq. \eqref{FP} and \eqref{P}. Hence, they can just be ignored.
In order to keep Eq. \eqref{FP} constant to the optimal value while
we minimize $p_{\text{succ}}^{\text{opt}}$, $e_{1 2^N}$ and $e_{2^N 1}$ must
be real and, thus, equal (see conditions 4. and 5.). Hence,
\begin{equation}
F^{\text{opt}}\cdot p_{\text{succ}}^{\text{opt}}=\frac{1}{2}\left[p_{\text{succ}}^{\text{opt}}+2e_{12^N}\sqrt{\epsilon(1-\epsilon)}^N\right].\label{FP2}
\end{equation}
From condition 5., $e^2_{1 2^N}$ is maximal when it reaches the maximum of $\text{min}(e_{11}e_{2^N 2^N},(1-e_{11})(1-e_{2^N
  2^N}))$, that is when $e_{11}e_{2^N 2^N}=(1-e_{11})(1-e_{2^N  2^N})$.
From this, it follows that $e_{11}+e_{2^N 2^N}=1$ and $e_{1
  2^N}=\sqrt{e_{11}e_{2^N 2^N}}$. Putting Eq. \eqref{FP2} equal to
$\frac{1}{2}\left[(1-\epsilon)^N+\epsilon^N\right]$, one gets the final form of $\Pi_B^{\text{succ,
    opt}}$, i.e.
\begin{equation}
\Pi_B^{\text{succ, opt}}=\left(
\begin{array}{ccccc}
\frac{(1-\epsilon)^N}{(1-\epsilon)^N+\epsilon^N}&0&\cdots&0&\frac{\sqrt{\epsilon (1-\epsilon)}^N}{(1-\epsilon)^N+\epsilon^N}\\
0&&&&0\\
\vdots&&\bigzero&&\vdots\\
0&&&&0\\
\frac{\sqrt{\epsilon (1-\epsilon)}^N}{(1-\epsilon)^N+\epsilon^N}&0&\cdots&0&\frac{\epsilon^N}{(1-\epsilon)^N+\epsilon^N}
\end{array}
\right).
\end{equation}
For this POVM the probability of success is
\begin{equation}
p_{\text{succ}}^{\text{opt}}=\frac{(1-\epsilon)^{2N}+\epsilon^{2N}}{(1-\epsilon)^N+\epsilon^N},\label{peps}
\end{equation}
and the fidelity is
\begin{equation}
F^{\text{opt}}=\frac{1}{2}\frac{\left[(1-\epsilon)^N+\epsilon^N\right]^2}{(1-\epsilon)^{2N}+\epsilon^{2N}}.\label{Feps}
\end{equation}
Note that $\Pi_B^{\text{succ}}=\mathbb{1}$ always retrieves the bound
of Eq. \eqref{eqF'},
with $p_{\text{succ}}=1$ and $F=\frac{1}{2}\left[(1-\epsilon)^N+\epsilon^N\right]$. Thus, the POVM
$\Pi_B^{\text{succ}}=w \Pi_B^{\text{succ, opt}}+(1-w)\mathbb{1}$,
i.e. an interpolation between the optimal measurement and the identity,
spans the threshold for all values of F and $p_{\text{succ}}$ that
optimize $F\cdot p_{\text{succ}}$.
\begin{figure}[h]
 \begin{center}
\includegraphics[width=250pt]{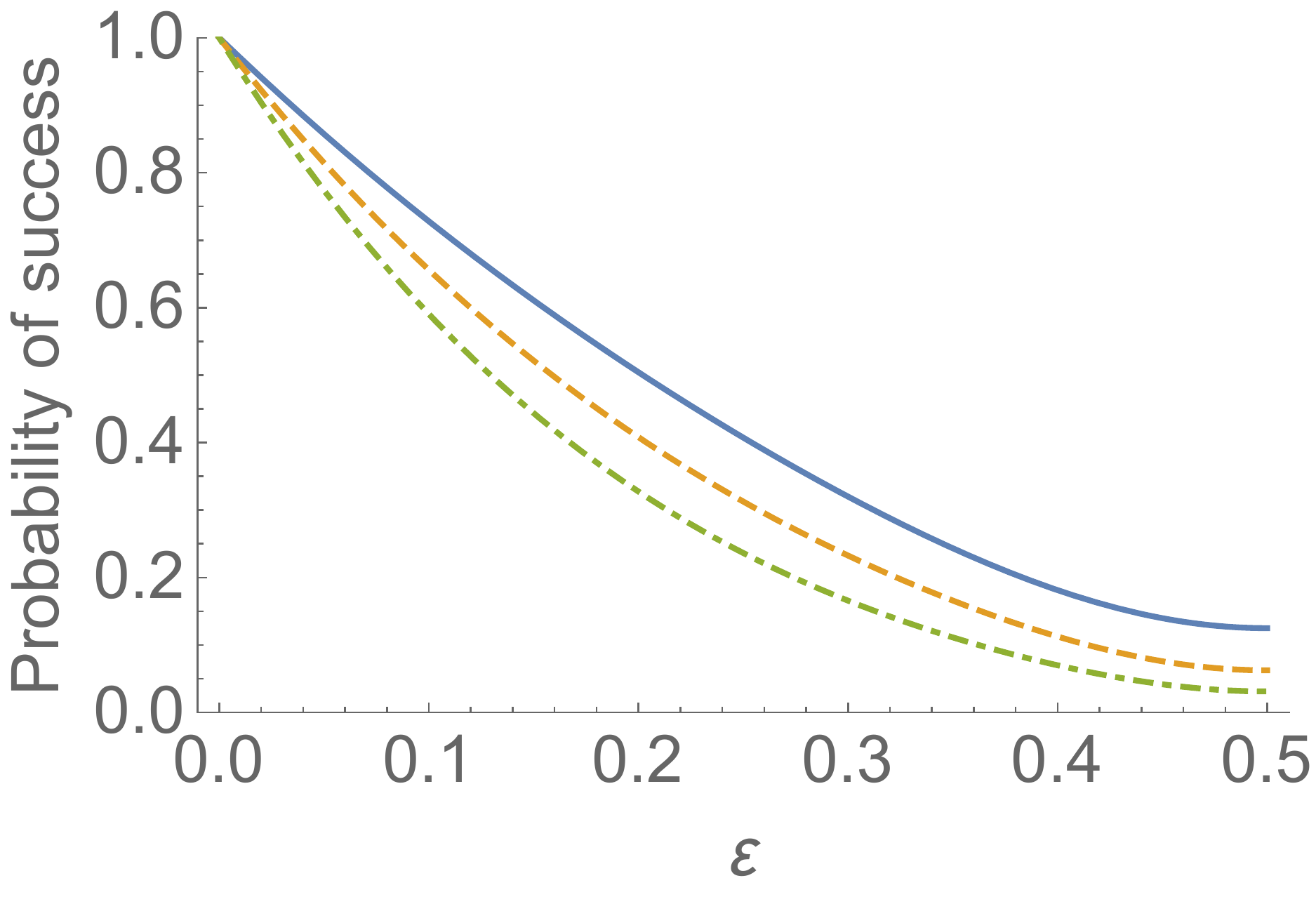}
 \end{center}
 \caption{Optimal success probability. The optimal success probability
   $p_{\text{succ}}^{\text{opt}}$ is plotted as a function of
   $\epsilon$ in the $[0,0.5]$ range for $N=3,4,$ and $5$ (full,
   dashed, and dotdashed curves, respectively). The
   function assumes values from 1 to $\frac{1}{2^N}$. The function
   goes from $1$ when no operation is performed over system B, to
   $\frac{1}{2^N}$, when a perfect GHZ projector is performed.\label{PEps}}
 \end{figure}
\begin{figure}[h]
 \begin{center}
\includegraphics[width=250pt]{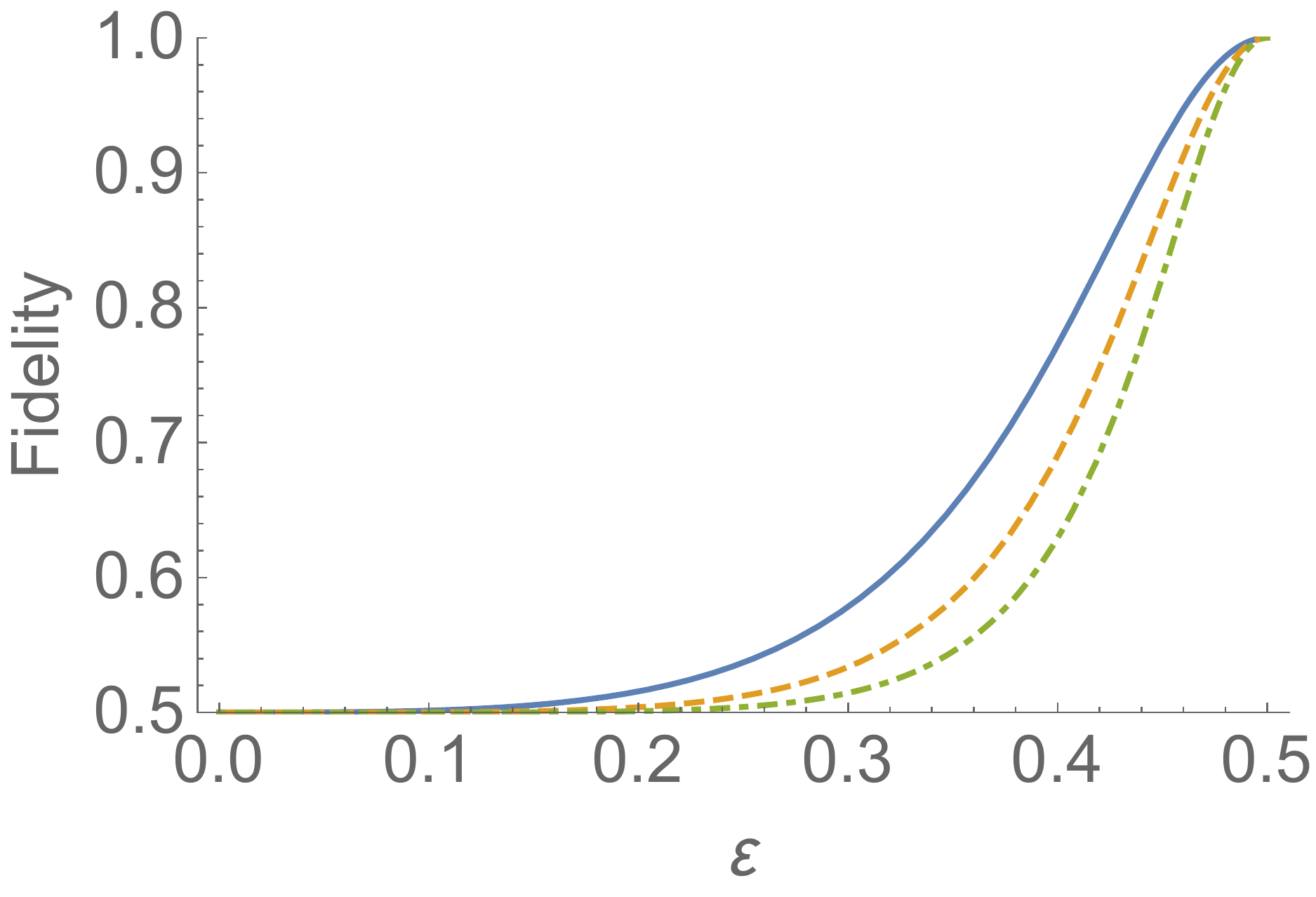}
 \end{center}
 \caption{Optimal fidelity. The optimal fidelity $F^{\text{opt}}$ is
   plotted as a function of $\epsilon$ for $N=3,4,$ and 5 (full,
   dashed, and dotdashed curves, respectively). Since the
   function is symmetrical respect to $\epsilon=0.5$, the plot is
   represented only in the $[0,0.5]$ range. The function goes from
   $0.5$, when no entanglement is generated, to 1, when the final
   state is the maximal entangled $\ket{\text{GHZ}}$.\label{FidEps}}
 \end{figure}
Eq. \eqref{peps} is the optimal success probability as a function of
$\epsilon$ and N when the bound of Eq. \eqref{eqF'} is attained and
F is maximal. Let's analyze the two Eqs. \eqref{peps} and
\eqref{Feps}. The two functions are plotted as a function of
$\epsilon$ for 3,4, and 5 nodes in Fig. \ref{PEps} and \ref{FidEps}, respectively.
Concerning the fidelity, it reaches the value 1 only for
$\epsilon=\frac{1}{2}$, i.e. for an initial maximally entangled state.
For other values of $\epsilon$, $F^{\text{opt}}$ is always smaller
than 1. The maximal success probability is obtained for
$\epsilon=0$. However, in this last case, the final state is
$\ket{0}\bra{0}_A^{\otimes N}$ and $F=\frac{1}{2}$, that is clearly an
uninteresting case since we are interested to high fidelity GHZ generation.
We can, then, conclude that the optimal case is $\epsilon=\frac{1}{2}$. $\Pi_B^{\text{succ, opt}}$ reduces to a
$\ket{\text{GHZ}}$ projector.

\section{Optical GHZ projector} \label{Sec:Implementation}
\begin{figure}[h]
 \begin{center}
 \includegraphics[width=190pt]{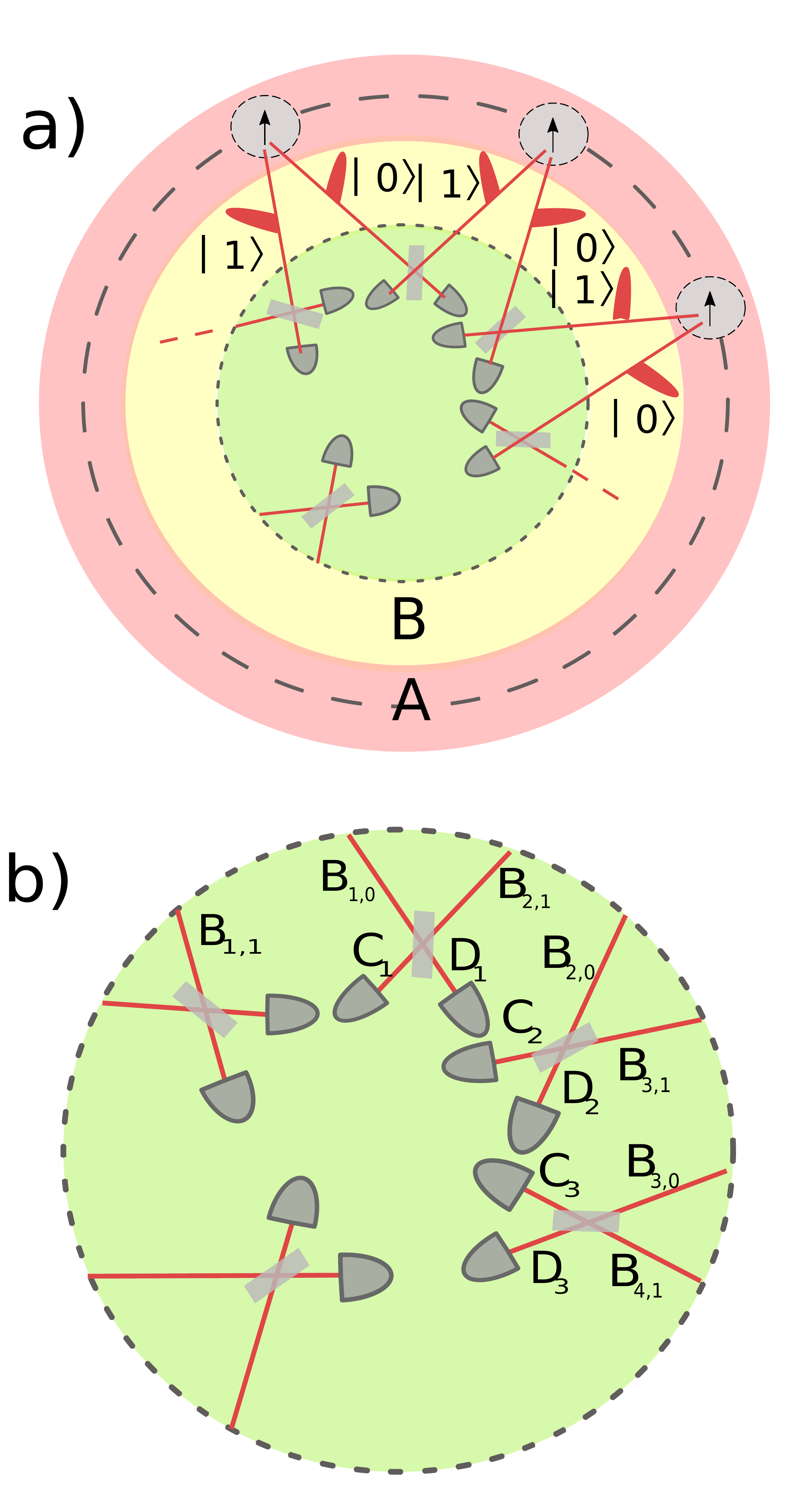}
 \end{center}
 \caption{Optical realization of a GHZ projector. a) Entire
   setup. Subsystem A (pink shell) is composed by N spin qubits on the external
   circle. The spin qubits are excited so that they generate the state
   $\ket{\Phi^+}_{AB}$ between them and N photonic qubits, composing subsystem B (yellow shell). Each photonic qubit is converted in an optical-path qubit through an optical switch and sent to the
   central station (green circle) where the projector is applied. b) Central
   station. Two optical-path modes coming from two neighbouring matter
   qubits impinge on the same beam splitter. A Bell measurement is
   performed on the modes $C$s and $D$s coming out from each beam splitter.\label{OpticalSys}}
 \end{figure}

\hspace{.5cm}
In this section we present a possible way of implementing a $\ket{\text{GHZ}}$ projector.
The envisioned setup is represented in Fig.\ref{OpticalSys}a)-b).
Subsystem A is the actual quantum network, composed by N nodes. Each
node is constituted by a quantum system with two long-lived spin
states \cite{Bernier13,Gao15,Hucul15,Delteil16},
here called $\ket{0}$ and $\ket{1}$, that can be independently excited
through optical pulses. As a consequence, each node is able to
generate a maximally entangled spin-photon pair, i.e.  $\frac{1}{\sqrt{2}}(\ket{0
  0}+\ket{1 1})_{A_jB_j}$, where $\ket{0}_{B_j}$ ($\ket{1}_{B_j}$) is a photon in
the 0 (1) mode.
The nodes interact between each other through the photonic
0-1 modes regrouped in subsystem B. The degree of freedom of modes in
subsystem B depends on the nature of the nodes. For example, for NV
centers \cite{Bernier13,Gao15}, and trapped ions \cite{Hucul15}, the photonic qubits can be encoded in
time-bin, and polarization, respectively.
The $j$th 0 and 1 modes are converted into two different spatial modes
$B_{j,0}$ and $B_{j,1}$, respectively. This can be done through an optical switch \cite{Lee10}. both directed to the central station
(Fig. \ref{OpticalSys}b)) where Bell measurements between modes $B_{j,0}$
and $B_{j+1,1}$ are performed.
In order to close the series of Bell measurements, a Bell measurement is performed between modes
$B_{1,1}$ and $B_{N,0}$.

\subsection{Proof\label{SubsectionProof}}
\hspace{.5cm}
 In this subsection, we prove that, depending on
 which set of detectors click, the setup in Fig. \ref{OpticalSys}
 allows to generate two GHZ-like states on subsystem A. Moreover, we derive the success
 probability as a function of the losses and we show that in the ideal
 case, each one of the final states has a probability of success
 $2^{-N}$, equivalent to the one one would get for a GHZ projector.
We split the proof into three steps. In step 1, we focus on one
successful combination of clicks among the $2^N$ successful
combinations. We prove that the event corresponding to this
particular combination of clicks happens with probability
$P_{\text{succ}}=\frac{2}{2^{2N}}\eta^N$, where $\eta$ is
the total probability that a photon does not get lost in the
transmission and detected by a detector, and that a GHZ-like state is,
thus, produced in A.
In step 2, we show that all the
other successful detector combinations projects system A into a GHZ-like state
with the same probability $P_{\text{succ}}$. In step 3, we show that the
set of GHZ-like states that are generated is composed by only two
states that differ by a relative phase. We calculate the total success
probability for each GHZ-like state and show that it is equal to
$P^{\text{Tot}}_{\text{succ}}=\left(\frac{\eta}{2}\right)^N$. For no
losses, $2^{-N}$ is the success probability that a GHZ projector $\ket{\text{GHZ}}\bra{\text{GHZ}}_B$ would project system A into a $\ket{\text{GHZ}}_A$ state.\\
The total state $\ket{\Phi^+}_{AB}$ generated between subsystems A and
B can be written in terms of creation
operators as
\begin{equation}
\ket{\Phi^+}_{AB}=\frac{1}{\sqrt{2}^N}\Pi_{j=1}^{N}\left(a^{\dagger}_{j,0}b^{\dagger}_{j,0}+a^{\dagger}_{j,1}b^{\dagger}_{j,1}\right)\ket{\mathbf{0}},
\end{equation}
where $\ket{\mathbf{0}}$ is the vacuum, $a^{\dagger}_{j,k}$ is the creation
operator of the $\ket{k}_{A_j}$ state on the jth spin qubit, and
$b^{\dagger}_{j,k}$ is the jth creation operator of the photonic mode $\ket{k}_{B_j}$.
 Each photonic mode $B_{j,k}$ is converted into a
sum of modes $C_m$ and $D_m$ when it impinges a beam splitter.
The equations that transform the operators
$b^{\dagger}_{j,k}$s are $b_{j,0}^{\dagger}=\frac{1}{\sqrt{2}}\left(i
  c_j^{\dagger}+d_j^{\dagger}\right)$ and $b_{j,1}^{\dagger}=\frac{1}{\sqrt{2}}\left(
  c_{j-1}^{\dagger}+id_{j-1}^{\dagger}\right)$.

Hence, the state becomes $\ket{\Psi}_{ACD}$, i.e.
\begin{equation}\begin{split}
\ket{\Psi}_{ACD}=\frac{1}{2^N}\Pi_{j=1}^{N}&\bigg(a^{\dagger}_{j,0}\left(i
    c_{j}^{\dagger}+d_{j}^{\dagger}\right)\\
&+a^{\dagger}_{j,1}\left(c_{j-1}^{\dagger}+i d_{j-1}^{\dagger}\right)\bigg) \ket{\mathbf{0}}.\\
\label{Psi1}
\end{split}\end{equation}

\textbf{Step 1}. Let's first focus on one single successful
combination of detections, that is when we get a
detection on all detectors on modes $C$s and none on modes $D$s. The
total detectors operator, composed by the no-click (click) operator
$D^D_{\text{nc}}$ ($D^C_{\text{c}}$) on modes $D$s ($C$s), is
$D^D_{\text{nc}}D^C_{\text{c}}=\Pi_{j=1}^N(1-\eta)^{d_j^{\dagger}d_j}\left[\mathbb{1}-(1-\eta)^{c_j^{\dagger}c_j}\right]$
\cite{Caprara15}. Let's analyze this operator in more details. Let's consider first the
click operator on a single mode $C_j$,
$D_c^{C_j}=\mathbb{1}-(1-\eta)^{c_j^{\dagger}c_j}=\sum_{n=1}^{+\infty}\left[1-(1-\eta)^n\right]\ket{n}\bra{n}_{C_j}$,
where $\ket{n}_{C_j}$ is a Fock state of n photons. The effect of the
operator $c_j^{\dagger}$ applied $l$ times on the right side of
$D_c^{C_j}$ is
\begin{equation}
D_c^{C_j}c_j^{\dagger l}=\sum_{n=l}^{+\infty}\left[1-(1-\eta)^n\right]\sqrt{\frac{n!}{(n-l)!}}\ket{n}\bra{n-l}.
\end{equation}
Note that in the previous equation if $l=0$ there is no
term in the sum with $\bra{0}_{C_j}$.
The previous remark implies that, in order to have a detection in
mode $C_j$, there must be at least a $c_j^{\dagger}$ in the detected
state, i.e. there must be at least one photon in mode $C_j$. Let's
consider, now, the operator
$D_{\text{nc}}^{D_j}=(1-\eta)^{d_j^{\dagger}d_j}=\sum_{n=0}^{+\infty}(1-\eta)^n\ket{n}\bra{n}_{D_j}$. In
this case, if there are no losses, the application of several
$d_j^{\dagger}$ gives a success only with no photon on mode $D_j$,
i.e. there are no $d_j^{\dagger}$ in the detected state. If losses occur, then there is a
non null probability of not having any detection in mode $D_j$, and, as
a consequence, a successful Bell measurement. Let's come back to the protocol.
 The N modes generate a photon
each. We need N detections, each one in one of the N $C$ modes. This
implies two things.
Firstly, the only states that have successful outcomes do not have
photons in any mode $D$, i.e. they do not have anyone of the operators
$d_j^{\dagger}$s. Secondly, in each mode $C$ there is only one
photon, i.e. in the final state each $c^{\dagger}_j$ appears only
once. We can, now, continue the calculations. We have
\begin{equation}\begin{split}
&\Tr_{CD}\left(\ket{\Psi}\bra{\Psi}_{ACD}D^D_{\text{nc}}D^C_{\text{c}}\right)=\\
\frac{1}{2^{2N}}\Tr_C\bigg[&\Pi_{j=1}^N\left(\mathbb{1}-(1-\eta)^{c_j^{\dagger}c_j}\right)\\
&\Pi_{j=1}^N(i a_{j,0}^{\dagger}c_j^{\dagger}+a_{j,1}^{\dagger}c_{j-1}^{\dagger})
  \ket{\mathbf{0}}\\
&\bra{\mathbf{0}}\Pi_{j=1}^N(-i
  a_{j,0}c_j+a_{j,1}c_{j-1}) \bigg]=\\
\left(\frac{\eta}{2^2}\right)^N&\left(\Pi_{j=1}^Ni a^{\dagger}_{j,0}+\Pi_{j=1}^Na_{j,1}^{\dagger}\right)  \ket{\mathbf{0}}\\
&\bra{\mathbf{0}}\left(\Pi_{j=1}^N(-i)
  a_{j,0}+\Pi_{j=1}^Na_{j,1}\right)=\\
\left(\frac{\eta}{2^2}\right)^N&\left(i^N\ket{0}^N+\ket{1}^N\right)\left((-i)^N\bra{0}^{\otimes
    N}+\bra{1}^{\otimes N}\right)_A=\\
2\left(\frac{\eta}{2^2}\right)^N&\ket{\text{GHZ-like}}\bra{\text{GHZ-like}}_A.
\end{split}\end{equation}
The prefactor in front of
$\ket{\text{GHZ-like}}\bra{\text{GHZ-like}}_A$ in the last passage is
the success probability of the set of Bell measurements $P_{\text{succ}}=\eta^N2^{1-2N}$.
Hence, the final state is
\begin{equation}
\ket{\text{GHZ-like}}_A=\frac{1}{\sqrt{2}}\left(i^N\ket{0}^{\otimes N}+\ket{1}^{\otimes N}\right)_A,
\end{equation}
that is a GHZ state except for a phase factor that can be easily
corrected.

\textbf{Step 2}. There are other
$2^N-1$ detector configurations that result in a success of the set of
 Bell measurements.\\
If we choose other click-no click configurations, we have to invert
the creation and annihilation operators for all modes where the success
Bell measurement combination has been changed,
i.e. $d_m^{(\dagger)}\leftrightarrow c_m^{(\dagger)}$.
It follows that the final states $\ket{\Psi_{\text{final}}}_A$ are of the same form, i.e.
\begin{equation}
\ket{\Psi_{\text{final}}}_A=\frac{1}{\sqrt{2}}\left(i^k\ket{0}^{\otimes N}+i^l\ket{1}^{\otimes N}\right)_A,
\end{equation}
but the relative phase between $\ket{0}^{\otimes N}$ and
$\ket{1}^{\otimes N}$ can change and depends on the specific combination.

 \textbf{Step 3}. Let's focus, now, on the calculation of the relative
 phase depending on the detector configuration. Each $c^{\dagger}_m$ gives an i
 phase term to
$\ket{0}^{\otimes N}$, while each $d^{\dagger}_m$ gives an i phase term to
$\ket{1}^{\otimes N}$. Therefore, the states generated by the
measurement are of the form
\begin{equation}
\frac{1}{\sqrt{2}}\left[i^{N-m} \ket{0}^{\otimes N}+i^m\ket{1}^{\otimes N}\right],\label{Psi3}
\end{equation}
with $m\in [0,N-1]$. The set of states given by Eq. \eqref{Psi3} is composed by only two states, up
to global phases, i.e.
\begin{center}
 $\frac{1}{\sqrt{2}}\left[i^{N} \ket{0}^{\otimes N}+\ket{1}^{\otimes N}\right]$,\hspace{.3cm}and\hspace{.3cm}
 $\frac{1}{\sqrt{2}}\left[i^{N} \ket{0}^{\otimes N}-\ket{1}^{\otimes N}\right]$.
\end{center}
Each state recurs the same number of times. Therefore, we have only two final
states, each one arising from $2^{N-1}$ configurations each. Per each final
GHZ-like state the total probability is, then
\begin{equation}
P^{\text{Tot}}_{\text{succ}}=2^{1-2N}\cdot 2^{N-1}\eta^N=2^{-N}\eta^N,\label{Ptot}
\end{equation}
that is the maximal probability of success that we can achieve. The
envisioned protocol generates, thus, two GHZ-like states, each one
with a $P^{\text{Tot}}_{\text{succ}}$ that in the ideal case is
$2^{-N}$. This probability corresponds to the success probability for
a GHZ projector.\\
With this the proof is complete. As a last remark, since depending on which detectors click there are
two different GHZ-like states, one can gain an extra factor 2 in the
total success probability for some applications.

\section{Performance} \label{Sec:Perf}
\hspace{.5cm}
In this section, we give some estimates of the performance of our
protocol.
\begin{figure}[h]
 \begin{center}
\includegraphics[width=250pt]{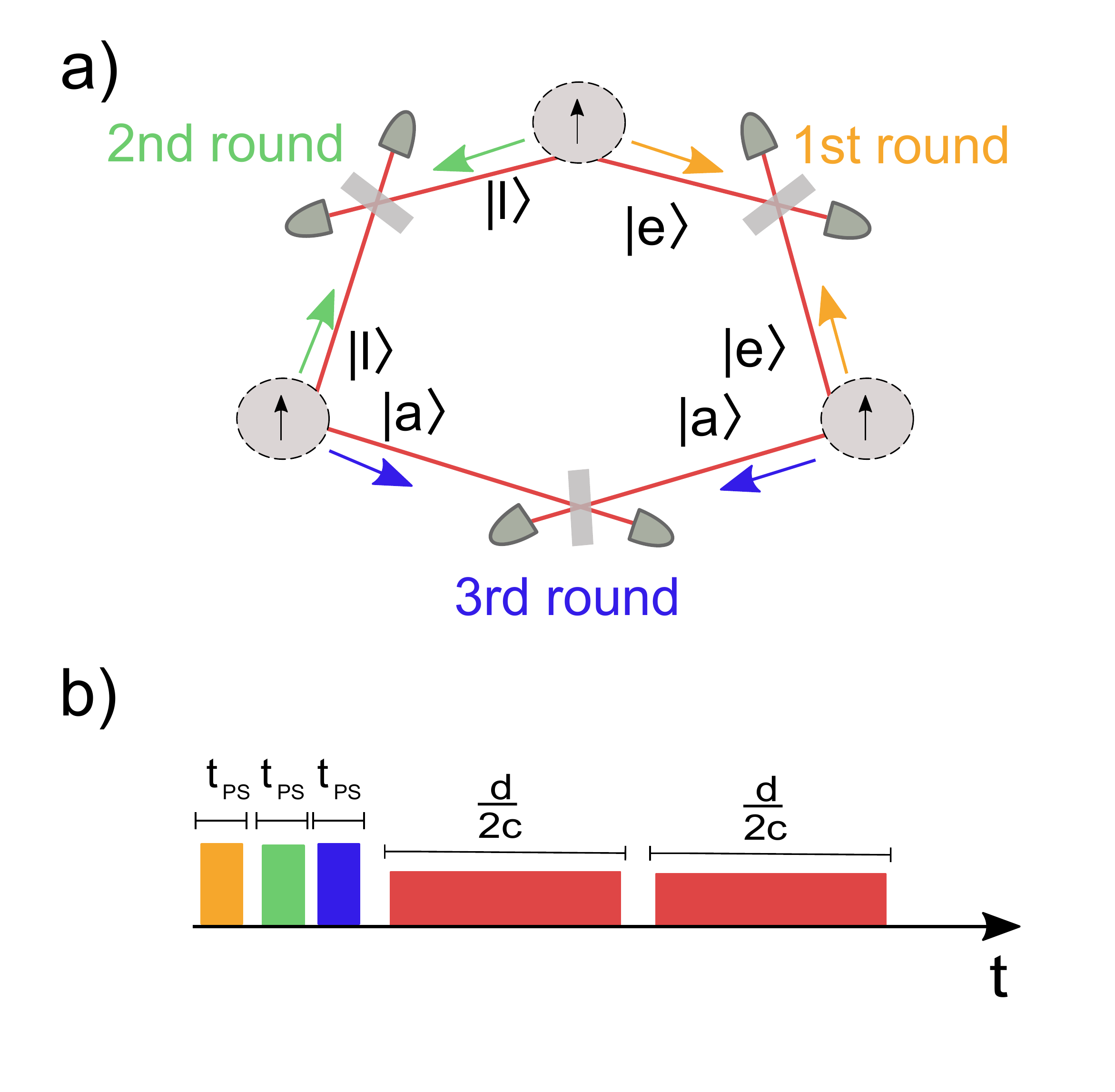}
 \end{center}
 \caption{Bell measurement synchronization. a) Three nodes protocol. In
   the figure it is represented the scheme in the case of three nodes
   as an example of an odd number of nodes. For an odd number of nodes
   it is necessary to perform three rounds for generating all the
   spin-photon pairs, early, late, and 'asynchronous'. b) Temporal scheme. In the picture it is
   represented the time line for an odd number of nodes. Three rounds
   are necessary for generating all the spin-photon
   pairs. Afterwards, the photons have to travel half d to reach the
   measurement station. Finally, the nodes have to wait till when the
   communication of the Bell measurement outcome comes back.\label{OddNumber}}
 \end{figure}
Since NV centers are promising candidates for quantum information
tasks \cite{Hensen15}, we consider values \cite{Dam17} of the involved quantities suitable for this
system. Let us remind you that for NV centers the photonic qubits can be
encoded in time-bin. The quantity that can be compared between different protocols
is the entanglement generation rate. The expression of
the generation rate $r_{\text{GHZ}}$ for GHZ states is
\begin{equation}
r_{\text{GHZ}}=\frac{P^{\text{Tot}}_{\text{succ}}}{t^{\text{Tot}}},
\end{equation}
where $P^{\text{Tot}}_{\text{succ}}$ is given by Eq. \eqref{Ptot}, and $t^{\text{Tot}}$ is the total time for each
protocol trial.
Hence, two factors influence the generation rate, namely the overall
transmission $\eta$ (see Eq. \eqref{Ptot}) and the time necessary to
perform each task involved in the protocol.
Let's focus on deriving $t^{\text{Tot}}$. It is given by the sum
of three quantities, i.e. the time necessary to generate all the
spin-photon pairs, the one necessary for the photons to travel half
the distance between two nodes, and the one necessary for
communicating to each node the outcome of the measurements.\\
For the sake of simplicity, for odd nodes the generated qubit pairs
are $\frac{1}{\sqrt{2}}(\ket{0e}+\ket{1l})_{AB}$, while for even nodes
the generated qubit pairs are
$\frac{1}{\sqrt{2}}(\ket{0l}+\ket{1e})_{AB}$, where $\ket{e}_B$
($\ket{l}_B$) is an early (late) photon.
The expression for the photon-spin generation time has a different expression
depending whether the number of nodes is odd or even. Indeed, in the
case of an even number, each early (late) mode is coupled with another
early (late) mode. On the contrary, in the case of an odd number of
nodes, in order to close the circle (see Fig. \ref{OpticalSys}), there
will be one branch, an ``asynchronous'' branch, where an early mode
would be coupled with a late mode
(see Fig. \ref{OddNumber}). This means that in the case of an even
number it is sufficient to consider only two rounds (early and late
mode) of photon generation per trial, while in the odd case we
consider a third round for the 'asynchronous branch'.
\begin{figure}[h]
 \begin{center}
\includegraphics[width=250pt]{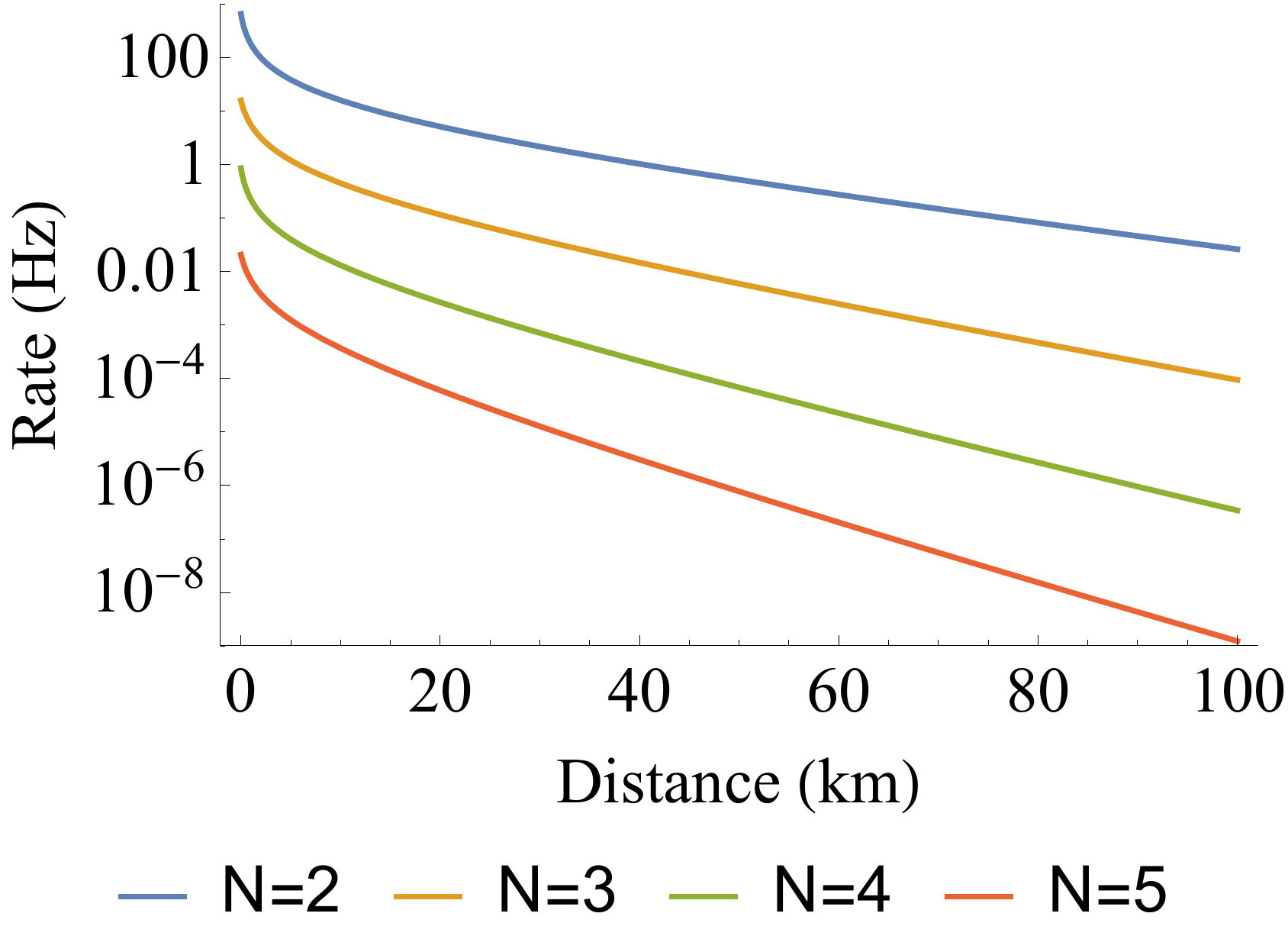}
 \end{center}
 \caption{GHZ generation rate $r_{\text{GHZ}}$. In the figure the generation rate for
   different values of $N$ (2,3,4, and 5) is plotted as a function of the distance
   $d$. The values of the experimental parameters are $L_0=20$ km, $\eta_{BS}=10^{-0.03}$ \cite{Thorlabs},
   $\eta_D=0.86$ \cite{Zadeh18}, $p_{\text{fc}}=0.3$, $p_{\text{out}}=0.3$, and
   $c=0.2\cdot 10^6$km/s. The total time $t^{\text{Tot}}$ for each
   attempt is given by Eq. \eqref{Ttote}-\eqref{Ttoto}.\label{PlotRate}}
 \end{figure}
 \begin{figure}
 \begin{center}
\includegraphics[width=250pt]{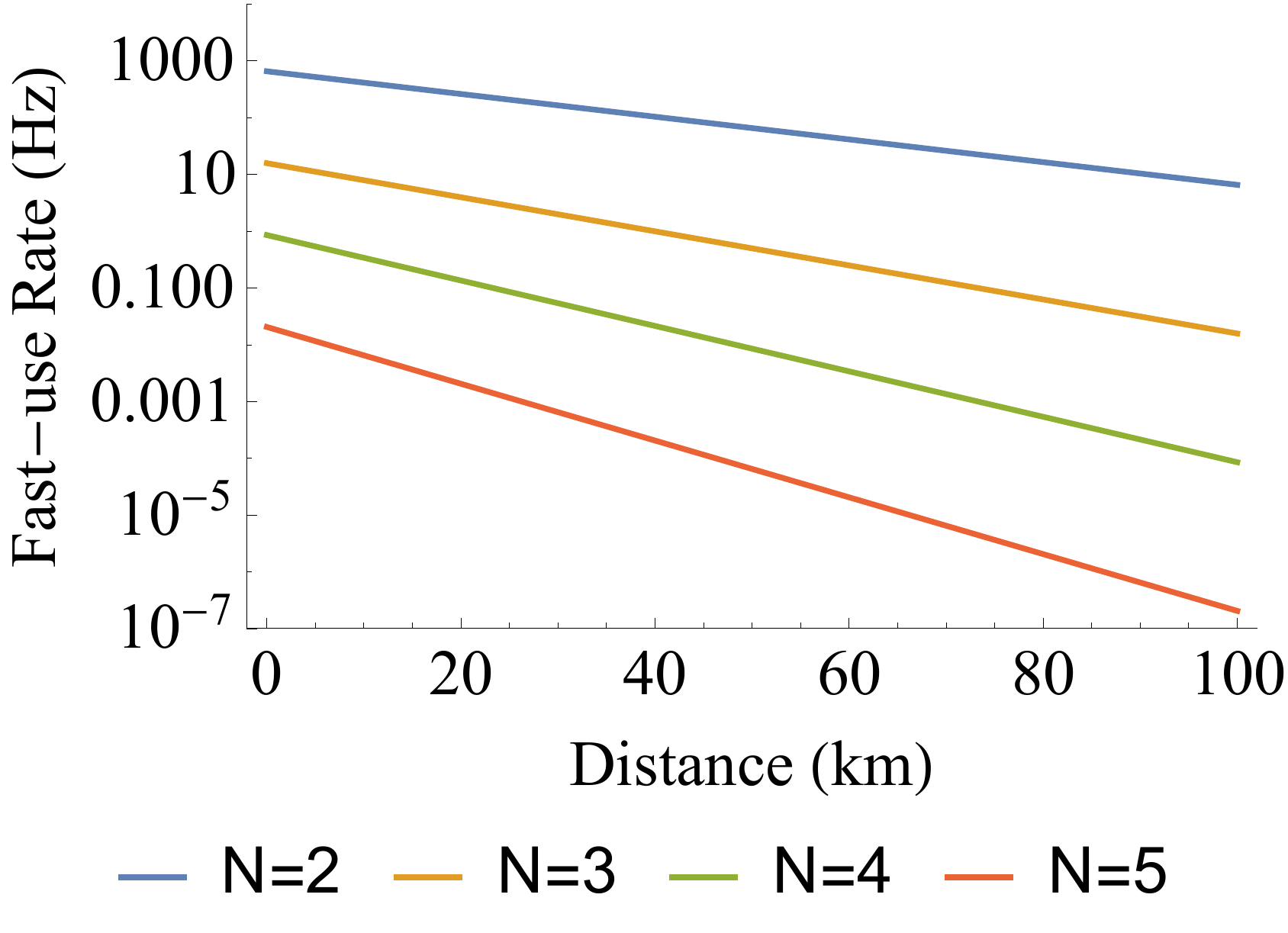}
 \end{center}
 \caption{Fast-use GHZ rate $r_{\text{GHZ}}$. In the figure the fast-use GHZ rate for
   different values of $N$ (2,3,4, and 5) is plotted as a function of the distance
   $d$. The values of the experimental parameters are $L_0=20$ km, $\eta_{BS}=10^{-0.03}$ \cite{Thorlabs},
   $\eta_D=0.86$ \cite{Zadeh18}, $p_{\text{fc}}=0.3$, $p_{\text{out}}=0.3$, and
   $c=0.2\cdot 10^6$km/s. The total
   time $t^{\text{Tot}}$ is given only by the spin-photon pair
   preparation time.\label{PlotUseRate}}
 \end{figure}
Let's define the time necessary to generate one
photon-spin pair as $t_{PS}$. For the sake of simplicity, we consider
that the distance $d$ between two neighbouring nodes is fixed. Given the
speed of light in an optical fibre $c$, the total generation time for an even number of
nodes is, then
\begin{equation}
t^{\text{Tot}}_{\text{even}}=2 t_{PS}+\frac{d}{2 c}+\frac{d}{2 c},\label{Ttote}
\end{equation}
while for an odd number is
\begin{equation}
 t^{\text{Tot}}_{\text{odd}}=3  t_{PS}+\frac{d}{2 c}+\frac{d}{2 c}.\label{Ttoto}
\end{equation}
Note that there is a factor 2, given by the fact
that the photons have to travel only half the distance between the
nodes in order to reach the Bell measurement station.
Note also that the second $\frac{d}{2c}$ factor is due to the
classical communication that has to be transmitted from the
measurement stations to the nodes.
Note also that there are some applications for which it is not
necessary to wait for the measurement and the arrival of the communication of the
outcome. In these cases the measurements on the nodes can be
realized straight away after that the photonic qubits have been sent
to the Bell measurement stations and the results kept (discarded) after
the communication of the success (failure) of the set of Bell
measurements. As a consequence, the total time $t^{\text{Tot}}$ can be
written for this case as just the spin-photon pairs preparation time. We call this rate the fast-use GHZ rate.
Concerning the overall transmission $\eta$, it is given by the formula \cite{Dam17}
\begin{equation}
\eta=\eta_{BS}\eta_D p_{\text{fc}}p_{\text{out}}10^{-\frac{\alpha d}{L_0}},
\end{equation}
where $L_0$ is the attenuation length of the fibres, $\alpha=0.2
\text{dB}/\text{km}$, $p_{\text{fc}}$ is the frequency conversion
efficiency, $p_{\text{out}}$ is the NV outcoupling efficiency, and $\eta_{BS}$
$\eta_D$ are the beam splitter and detector efficiency.
In Figs. \ref{PlotRate} and \ref{PlotUseRate}, there are the results for the rate as a function of the
distance between two nodes for $N=2,3,4$ and $5$.
In Fig. \ref{PlotRate} there are the plots for the case $t^{\text{Tot}}$
is given by Eq. \eqref{Ttote}-\eqref{Ttoto}, while in
Fig. \ref{PlotUseRate} the plots are made for $t^{\text{Tot}}$ only equal
to the spin-photon preparation time.
As one might expect from Eq. \eqref{Ptot}, the curves decrease of one
term $\frac{\eta}{2}$ per each added node.
However, while in Fig. \ref{PlotRate} the curves are proportional to
$d^{-1}10^{-\frac{\alpha N d}{L_0}}$, in Fig. \ref{PlotUseRate} they are
proportional only to the exponential term $10^{-\frac{\alpha N
    d}{L_0}}$. This results in an improvement of two orders of
magnitude more in the second case for $d=100$ km.

\section{Conclusion}
\hspace{.5cm}
The protocol that we have presented here is an adaptation for matter
systems and an arbitrary number of nodes of a protocol \cite{Zukowski95} meant for fully optical
systems and only three parties. We consider the protocol interesting
for several reasons. Firstly, it is a natural extension for N nodes of the well
known Barrett-Kok scheme \cite{Barrett05,Lim05} and so it is
particularly suited for achieving high fidelities. Secondly, we have
proven that in the ideal case, i.e. in the case of no loss, the success probability is optimal. This
is, indeed, quite surprising since the scheme is based only on linear
optics. Nonetheless, there are some aspects that deserve some
attention. Indeed, in a real scenario all the causes of loss and noise have to be taken into account. Unfortunately, optimizing a
scheme in a real scenario, where such kind of processes are involved, is a challenging task. However, it seems to
us that the required resources and causes of decoherence and
depolarization in our case are minimal. Thus, the protocol is likely
optimal also when losses and noise occur.
Our scheme presents two intrinsic drawbacks, in that, it can only be
implemented between nearby nodes and the performance showed in the
previous section is quite low. It is, then, of interest to
evaluate other protocols, that combine distillation procedures with
Bell measurements.
In this case the parameter of reference would be the generation rate
and not anymore the success probability. However, all these protocols would intrinsically be
affected by decoherence that would inevitably lower the
fidelity. They are not, then, competitive in the high fidelity regime
that we have explored in this article.
It is still interesting to investigate if there exists procedures both
for nearby and distant nodes that allow to appealing trade-offs
between generation rate and final fidelity.

\section{Supplemental Material:Numerical Optimization\label{3}}
\hspace{.5cm}
In the main text, we have analitically optimized $F\cdot
p_{\text{succ}}$ and showed how to experimentally retrieve this value.\\
In this supplemental material, we explain how to perform numerical optimizations
over POVMs in order to optimize $F\cdot p_{\text{succ}}$ for arbitrary input states.
One can retrieve the previous expressions for $F\cdot
p_{\text{succ}}$ and $p_{\text{succ}}$ in terms of a map $\Lambda$
acting on system B.
The expression for $F\cdot
p_{\text{succ}}$ becomes
\begin{equation}\begin{split}
F\cdot p_{\text{succ}}=\Tr((\ket{\text{GHZ}}\bra{\text{GHZ}}_A\otimes\left(\ket{\text{GHZ}}\bra{\text{GHZ}}\cdot
  \Lambda\right)_B)&\\
\cdot\ket{\Psi_{\text{in}}}\bra{\Psi_{\text{in}}}_{AB})&,
\end{split}\end{equation}
where we have substituted
$\Pi^{\text{succ}}_B=\left(\ket{\text{GHZ}}\bra{\text{GHZ}}\cdot
  \Lambda\right)_B$, $\Lambda_B$ being an arbitrary map.
In a similar way the success probability $p_{\text{succ}}$ takes
the following form
\begin{equation}
 p_{\text{succ}}=\Tr\left[\left( \mathbb{1}_A\otimes
                (\ket{\text{GHZ}}\bra{\text{GHZ}}\cdot \Lambda)_B\right)\cdot\ket{\Psi_{\text{in}}}\bra{\Psi_{\text{in}}}_{AB}\right].
\end{equation}
Our goal is to find the optimal $\Lambda_B$, subject to a
fixed $p_{\text{succ}}$, such that the product $F\cdot p_{\text{succ}}$ is maximal.

\subsection{Choi-Jamiolkowski Isomorphism}
\hspace{.5cm}
One can realize the previous optimization using the Choi-Jamiolkowski
isomorphism. Let's assume to have two systems $S$ and $S'$ of the same
dimension $|S|$.
 Given the positive map $\Lambda_{S'}$, acting on $S'$, the Choi's theorem states that the matrix
\begin{equation}
\tau_{\text{SS'}}=\mathbb{1}_S\otimes \Lambda_{S'}(\ket{\Phi^+}_{SS'}),
\end{equation}
where $\ket{\Phi^+}_{SS'}=\frac{1}{\sqrt{|S|}}\sum_{m=1}^{|S|}\ket{mm}_{ss'}$ is a maximally entangled state between systems S and $S'$, has the properties
\begin{enumerate}
\item $\tau_{SS'}\ge 0$,
\item $\Tr(\tau_{SS'})=1$, and
\item $\tau_S=\Tr_{S'}(\tau_{SS'})=\frac{\mathbb{1}_S}{|S|}$.
\end{enumerate}
Given the above-listed first two properties, $\tau_{SS'}$ is a density matrix
and it is called Jamiolkowski state.\\

\subsection{Initial Maximally Entangled State}
\hspace{.5cm}
In the case of an initial maximally entangled state, for example
$\ket{\Psi_{\text{in}}}_{AB}=\ket{\Phi^+}_{AB}=\bigotimes_{j=1}^N\frac{1}{\sqrt{2}}(\ket{00}+\ket{11})_{A_j
  B_j}$, the map $\mathbb{1}_A\otimes\Lambda_{B}$ applied
to $\ket{\Psi_{\text{in}}}_{AB} $ is a Jamiolkowski state,
i.e. $\mathbb{1}_A\otimes\Lambda_B(\ket{\Phi^+}_{AB})=\tau_{AB}$
is a state. The quantities $F\cdot p_{\text{succ}}$ and
$p_{\text{succ}}$ can be rewritten in terms of the Jamiolkowski state,
i.e.
\begin{equation}
F\cdot p_{\text{succ}}=\Tr\left[ \left(\ket{\text{GHZ}}\bra{\text{GHZ}}_A\otimes
  \ket{\text{GHZ}}\bra{\text{GHZ}}_B\right)\cdot\tau_{AB}\right],
\end{equation}
and
\begin{equation}
p_{\text{succ}}=\Tr\left[ \left(\mathbb{1}_A\otimes
  \ket{\text{GHZ}}\bra{\text{GHZ}}_B\right)\cdot\tau_{AB}\right].
\end{equation}
 Hence, the optimization becomes:
\fbox{\parbox{0.47\textwidth}{Max $F\cdot p_{\text{succ}}$ s.t.
\begin{enumerate}
\item $\tau_{AB}\ge 0$,
\item $\Tr(\tau_{AB})=1$,
\item $\tilde{\tau}_{A}=\Tr_B(\tau_{AB})=\frac{\mathbb{1}_A}{2^N}$,
\item $p_{\text{succ}}$ is fixed.
\end{enumerate}
}%
}
The first three conditions are equivalent to the ones of subsection
A., while the last is necessary for deriving $F\cdot p_{\text{succ}}$
as a function of $p_{\text{succ}}$.
\begin{figure}[h]
 \begin{center}
\includegraphics[width=250pt]{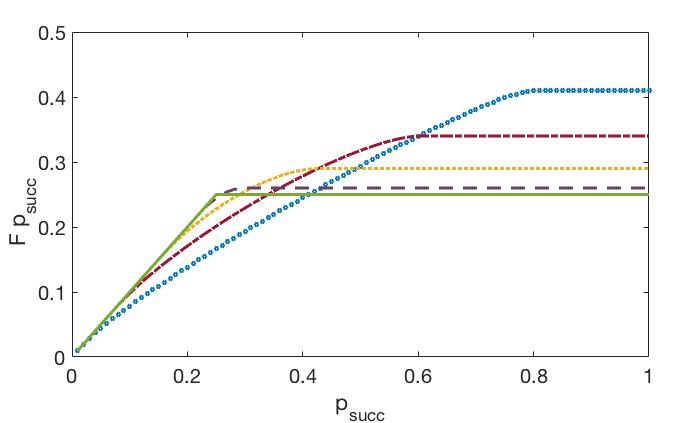}
 \end{center}
 \caption{Numerical optimization of the product $F\cdot
   p_{\text{succ}}$ as a function of $p_{\text{succ}}$ for $N=2$. The
   curves represent different values of $\epsilon=0.5,0.4,0.3,0.2,$
   and $0.1$ (Full, dashed, dotted, dotdashed, and marble).}
 \end{figure}

\subsection{Initial Non-Maximally Entangled State}
\hspace{.5cm}
Consider now the case when the initial state is non-maximally
entangled,
for example $\ket{\Psi_{\text{in}}}_{AB}=\bigotimes_{j=1}^N(\sqrt{1-\epsilon}\ket{00}+\sqrt{\epsilon}\ket{11})_{A_j
  B_j}$. Let's put system $S$ ($S'$) equal to the initial (final)
system AB. One can apply the
Choi-Jamiolkowski isomorphism to $F\cdot
p_{\text{succ}}$ and $p_{\text{succ}}$. Indeed, we have
\begin{equation}\begin{split}
F\cdot
p_{\text{succ}}=2^{2N}\Tr&\bigg[\ket{\Psi_{\text{in}}}\bra{\Psi_{\text{in}}}_{AB}^{\text{in}}\otimes \big(\ket{\text{GHZ}}\bra{\text{GHZ}}_A\\
&\otimes
    \ket{\text{GHZ}}\bra{\text{GHZ}}_B \big)^{\text{fin}}\cdot\tilde{\tau}_{AB}\bigg],
\end{split}\end{equation}
and
 \begin{figure}[h]
 \begin{center}
\includegraphics[width=250pt]{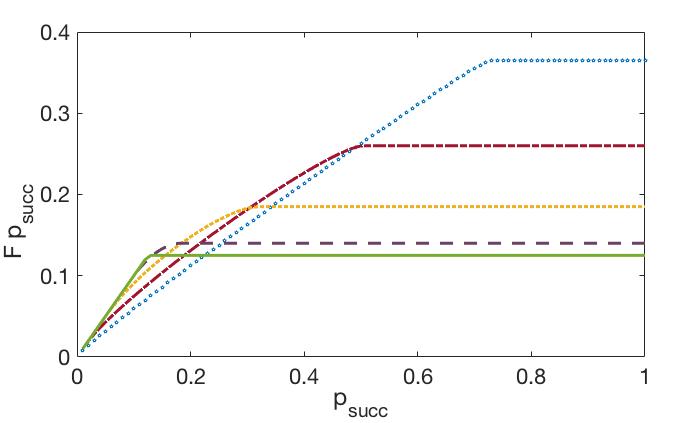}
 \end{center}
 \caption{Numerical optimization of the product $F\cdot
   p_{\text{succ}}$ as a function of $p_{\text{succ}}$ for $N=3$. The
   curves represent different values of $\epsilon=0.5,0.4,0.3,0.2,$
   and $0.1$ (Full, dashed, dotted, dotdashed, and marble).}
 \end{figure}
\begin{equation}
p_{\text{succ}}=2^{2N}\Tr[ \ket{\Psi_{\text{in}}}\bra{\Psi_{\text{in}}}^{\text{in}}_{AB}\otimes \left( \mathbb{1}_A\otimes
                 \ket{\text{GHZ}}\bra{\text{GHZ}}_B\right)^{\text{fin}} \tilde{\tau}_{AB}],
\end{equation}
where $\tilde{\tau}_{AB}=\mathbb{1}^{\text{in}}_{AB}\otimes
\tau^{\text{fin}}_{AB}$. Here, $2^{2N}$ is the dimension of one of the
two subsystems initial and final.
 Hence, we want to perform the following optimization:\\
\fbox{\parbox{0.47\textwidth}{Max $F\cdot p_{\text{succ}}$ s.t.
\begin{enumerate}
\item $\tau^{\text{fin}}_{AB}\ge 0$,
\item $\Tr(\tau^{\text{fin}}_{AB})=1$,
\item
$\tilde{\tau}^{\text{fin}}_{A}=\Tr_{B}(\tau^{\text{fin}}_{AB})=\frac{\mathbb{1}_{A}}{2^N}$,
\item $p_{\text{succ}}$ is fixed.
\end{enumerate}
}%
}
The above explained optimization has been performed for three nodes,
providing results in perfect agreement with the analytical upperbounds
derived in the main text.

\end{document}